\documentclass[twocolumn,nofootinbib,amsmath,amssymb,a4paper]{revtex4}

\usepackage{graphicx}
\usepackage{dcolumn}
\usepackage{bm}

\RequirePackage{xspace}

\usepackage{relsize}
\def\babar{\mbox{\slshape B\kern-0.1em{\smaller A}\kern-0.1em
    B\kern-0.1em{\smaller A\kern-0.2em R}}}

\def\epem       {\ensuremath{e^+e^-}\xspace}

\def\c     {\ensuremath{c}\xspace}

\def\b     {\ensuremath{b}\xspace}

\def\piz   {\ensuremath{\pi^0}\xspace}

\def\pip   {\ensuremath{\pi^+}\xspace}
\def\pim   {\ensuremath{\pi^-}\xspace}

\def\Kbar  {\kern 0.2em\overline{\kern -0.2em K}{}\xspace}

\def\Kz    {\ensuremath{K^0}\xspace}
\def\Kzb   {\ensuremath{\Kbar^0}\xspace}
\def\KzKzb {\ensuremath{\Kz \kern -0.16em \Kzb}\xspace}
\def\Kp    {\ensuremath{K^+}\xspace}
\def\Km    {\ensuremath{K^-}\xspace}

\def\KpKm  {\ensuremath{\Kp \kern -0.16em \Km}\xspace}
\def\KS    {\ensuremath{K^0_{\scriptscriptstyle S}}\xspace}

\def\Kstarzb {\ensuremath{\Kbar^{*0}}\xspace}

\def\Dbar    {\kern 0.2em\overline{\kern -0.2em D}{}\xspace}

\def\Dz      {\ensuremath{D^0}\xspace}
\def\Dzb     {\ensuremath{\Dbar^0}\xspace}
\def\DzDzb   {\ensuremath{\Dz {\kern -0.16em \Dzb}}\xspace}
\def\Dp      {\ensuremath{D^+}\xspace}
\def\Dm      {\ensuremath{D^-}\xspace}

\def\DpDm    {\ensuremath{\Dp {\kern -0.16em \Dm}}\xspace}
\def\Dstar   {\ensuremath{D^*}\xspace}

\def\Dstarz  {\ensuremath{D^{*0}}\xspace}

\def\Dstarmp {\ensuremath{D^{*\mp}}\xspace}
\def\Ds      {\ensuremath{D^+_s}\xspace}

\def\B       {\ensuremath{B}\xspace}
\def\Bbar    {\kern 0.18em\overline{\kern -0.18em B}{}\xspace}

\def\BB      {\ensuremath{B\Bbar}\xspace} 
\def\Bz      {\ensuremath{B^0}\xspace}
\def\Bzb     {\ensuremath{\Bbar^0}\xspace}
\def\BzBzb   {\ensuremath{\Bz {\kern -0.16em \Bzb}}\xspace}
\def\Bu      {\ensuremath{B^+}\xspace}
\def\Bub     {\ensuremath{B^-}\xspace}

\def\BpBm    {\ensuremath{\Bu {\kern -0.16em \Bub}}\xspace}

\def\BorBbar    {\kern 0.18em\optbar{\kern -0.18em B}{}\xspace}
\def\DorDbar    {\kern 0.18em\optbar{\kern -0.18em D}{}\xspace}
\def\KorKbar    {\kern 0.18em\optbar{\kern -0.18em K}{}\xspace}

\mathchardef\Upsilon="7107
\def\Y#1S{\ensuremath{\Upsilon{(#1S)}}\xspace}

\def\FourS {\Y4S}

\mathchardef\Deltares="7101
\mathchardef\Xi="7104
\mathchardef\Lambda="7103
\mathchardef\Sigma="7106
\mathchardef\Omega="710A

\def\Deltabar{\kern 0.25em\overline{\kern -0.25em \Deltares}{}\xspace}
\def\Lbar{\kern 0.2em\overline{\kern -0.2em\Lambda\kern 0.05em}\kern-0.05em{}\xspace}
\def\Sigbar{\kern 0.2em\overline{\kern -0.2em \Sigma}{}\xspace}
\def\Xibar{\kern 0.2em\overline{\kern -0.2em \Xi}{}\xspace}
\def\Obar{\kern 0.2em\overline{\kern -0.2em \Omega}{}\xspace}
\def\Nbar{\kern 0.2em\overline{\kern -0.2em N}{}\xspace}
\def\Xb{\kern 0.2em\overline{\kern -0.2em X}{}\xspace}

\def\BR         {{\ensuremath{\cal B}\xspace}}

\def\mes        {\mbox{$m_{\rm ES}$}\xspace}

\def\DeltaE     {\mbox{$\Delta E$}\xspace}

\newcommand{\tev}{\ensuremath{\mathrm{\,Te\kern -0.1em V}}\xspace}
\newcommand{\gev}{\ensuremath{\mathrm{\,Ge\kern -0.1em V}}\xspace}
\newcommand{\mev}{\ensuremath{\mathrm{\,Me\kern -0.1em V}}\xspace}
\newcommand{\kev}{\ensuremath{\mathrm{\,ke\kern -0.1em V}}\xspace}
\newcommand{\ev}{\ensuremath{\mathrm{\,e\kern -0.1em V}}\xspace}
\newcommand{\gevc}{\ensuremath{{\mathrm{\,Ge\kern -0.1em V\!/}c}}\xspace}
\newcommand{\mevc}{\ensuremath{{\mathrm{\,Me\kern -0.1em V\!/}c}}\xspace}
\newcommand{\gevcc}{\ensuremath{{\mathrm{\,Ge\kern -0.1em V\!/}c^2}}\xspace}
\newcommand{\mevcc}{\ensuremath{{\mathrm{\,Me\kern -0.1em V\!/}c^2}}\xspace}

\def\mus  {\ensuremath{\rm \,\mus}\xspace}

\def\mus        {\ensuremath{\,\mu{\rm s}}\xspace}    

\def\ra                 {\ensuremath{\rightarrow}\xspace}
\def\to                 {\ensuremath{\rightarrow}\xspace}

\def\pep2{PEP-II}

\def\gsim{{~\raise.15em\hbox{$>$}\kern-.85em
          \lower.35em\hbox{$\sim$}~}\xspace}
\def\lsim{{~\raise.15em\hbox{$<$}\kern-.85em
          \lower.35em\hbox{$\sim$}~}\xspace}

\def\CP                {\ensuremath{C\!P}\xspace}

\def\P       {\ensuremath{P}\xspace}

\xspace

\newcommand{\jprlBase}       {Phys.\ Rev.\ Lett.\xspace}
\newcommand{\jprBase}        {Phys.\ Rev.\xspace}
\newcommand{\jplBase}        {Phys.\ Lett.\xspace}

\newcommand{\plb}       [1]  {\jplBase\ B~{\bf #1}}

\newcommand{\jprl}      [1]  {\jprlBase\ {\bf #1}}
\newcommand{\jprd}      [1]  {\jprBase\ D~{\bf #1}}

\def\jetset74   {\mbox{\tt Jetset \hspace{-0.5em}7.\hspace{-0.2em}4}\xspace}

\def\dt{\Delta t}
\def\Brec{B_{\rm rec}}
\def\Btag{B_{\rm tag}}

\def\mmiss{m_{\rm miss}}

\def\btodstpipm{\Bz \rightarrow \Dstarmp\pi^\pm}

\def\r{{r^{(*)}}}

\begin{document}

\title{\babar\ status and prospects for \CP asymmetry measurements: $\sin(2\beta+\gamma)$}

\author{S. Ganzhur}
 \email{ganzhur@cea.fr}
\affiliation{%
DSM/Dapnia, CEA/Saclay, F-91191 Gif-sur-Yvette, France\\
}%

\begin{abstract}
The recent experimental results on \CP violation related to the angles of the
Cabibbo-Kobayashi-Maskawa (CKM) unitarity triangle $2\beta+\gamma$
are summarized in these proceedings. These results are obtained with approximately $232$ 
million $\Y4S{\to}\BB$ events collected with the \babar\ detector at the PEP-II asymmetric-energy 
$B$-factory at SLAC.  Using the  measurements on time-dependent \CP\ asymmetries in 
$\Bz{\to}D^{(*)\mp}\pi^{\pm}$ and $\Bz{\to}D^{\mp}\rho^{\pm}$ decays and theoretical assumptions, 
one finds $|\sin(2\beta+\gamma)|\!>\!0.64\ (0.40)$ at $68\%\ (90\%)$ confidence level.
The perspectives of $\sin(2\beta+\gamma)$ measurement with 
$\Bzb\to D^{(*)0}\bar{K}^{(*)0}$ and $\Bz \rightarrow D^{(*)\mp} a_{0(2)}^{\pm}$ decay 
channels are also discussed.
\end{abstract}

\maketitle

\section*{Introductoin}
A crucial part of the \CP violation  program in \B-factories is the measurement of the angle
$
\gamma (\phi_3) = \arg{\left(- V^{}_{ud} V_{ub}^\ast/ V^{}_{cd} V_{cb}^\ast\right)} 
$
of the unitary triangle related to the CKM matrix~\cite{bib:KM}.  
Decays of $B_d$ mesons that allows one to constraint the 
CKM angle $2\beta+\gamma$, have either small \CP asymmetry ($\B\to D^{(*)}\pi$) 
or small branching fractions ($\B\to D^{(*)}K^{(*)}$). 
This makes the \CP violation effect hard to measure. Furthermore,
due to presence of  two hadronic parameters in the observables ($r$ and $\delta$, the amplitude ratio and 
the strong phase difference between two amplitudes) it is difficult to cleanly extract 
the weak phase information, although approaches based on SU(3) symmetry exists.

\section{The \babar\ detector and dataset}
\label{sec:babar}

The data used in the presented analyzes were recorded with the \babar\
detector at the \pep2\ asymmetric-energy storage rings, and consist of 211~fb$^{-1}$
collected on the $\Upsilon(4{\rm S})$ resonance (on-resonance
sample), and 21~fb$^{-1}$ collected at an $\epem$ center-of-mass (CM)
energy approximately 40~\mev below the resonance peak
(off-resonance sample). This corresponds to approximately $232$ million $\Y4S{\to}\BB$ recorded events. 

The \babar\ detector is described in detail in Ref.~\cite{ref:babar}.

\section{\CP asymmetry in $\Bz \rightarrow D^{(*)\mp} \pi^{\pm}/\rho^\pm$ decays}
The decay modes $\Bz \rightarrow D^{(*)\mp} \pi^{\pm}$ have been
proposed  to measure
$\sin(2\beta+\gamma)$~\cite{ref:sin2bg_th}.
In the Standard Model the decays
$\Bz \to D^{(*)+} \pi^-$ and $\Bzb \to D^{(*)+} \pi^-$
proceed through the $\overline{b} \rightarrow \overline{u}  c  d $ and
$\b\to c$ amplitudes $A_u$ and $A_c$, respectively.
The relative weak phase between these two amplitudes
is $\gamma$. When combined with $\Bz \Bzb$ mixing, this yields a weak phase
difference of $2\beta+\gamma$ between the interfering amplitudes.

The decay rate distribution for 
$B \to D^{(*)\pm}\pi^\mp$ is
\begin{eqnarray}
\P^\pm_\eta(\dt)
&=& {e^{-|\dt|/\tau} \over 4\tau} \times       
\left[ 1 \mp S^\zeta \sin(\Delta m \dt) \right.  \nonumber\\  
& & \left.\mp \eta C \cos(\Delta m \dt) \right], 
\label{eq:pure-dt-pdf-B}
\end{eqnarray}
where  $\tau$ is the $\Bz$ lifetime averaged over the two mass eigenstates,
$\Delta m$ is the $\Bz-\Bzb$ mixing frequency, and $\dt$
is the difference between the time
of the $B\to{D^{(*)\pm}}\pi^\mp$ ($\Brec$)
decay and the decay of the other
$B$ ($\Btag$) in the event. The
upper (lower) sign in Eq.~\ref{eq:pure-dt-pdf-B}
indicates the flavor of the $\Btag$ as a $\Bz$ ($\Bzb$),
while $\eta = +1$ ($-1$) and $\zeta = +$ ($-$) for
the $\Brec$ final state ${D^{(*)-}}\pi^+$ (${D^{(*)+}}\pi^-$).
The parameters $C$ and $S^\pm$ are given by
\begin{equation}
C \equiv {1 - r^2 \over 1 + r^2}\, , \ \ \ \
S^\pm \equiv {2 r \over 1 + r^2}\, \sin(2 \beta + \gamma \pm \delta).
\end{equation}
Here $\delta$ is the strong phase difference
between $A_u$  and  $A_c$ and $r \equiv |A_u / A_c|$.
Since $A_u$ is doubly CKM-suppressed with respect
to $A_c$, one expects $r$ to be small of order 2\%.
Due to the small value of $r$, large data samples
are required for a statistically significant measurement of $S$. 

Since the expected \CP\ asymmetry in the selected $B$ 
decays is small, this measurement is sensitive to the interference between the 
$b{\to}u$ and  $b{\to}c$ amplitudes in the decay of $\Btag$.
To account for this ``tagside interference'', we use a parametrization 
which is described in Ref.~\cite{ref:abc}. The $S^\pm$ coefficient are replaced with three 
others 
\begin{eqnarray} 
  a &=& 2 r \sin(2\beta+\gamma) \cos \delta \,          \nonumber \\
  b &=& 2 r' \sin(2\beta+\gamma) \cos \delta' \,        \\
  c &=& 2 \cos(2\beta+\gamma) (r\sin \delta-r'\sin \delta') \nonumber 
\end{eqnarray}
For each tagging category, independent of the decay mode $\{D\pi, D^{*}\pi, D\rho\}$,  
the tagside interference is parametrized in terms of the
effective parameters $r'$ and $\delta'$. One notes, $r'=0$ for the lepton tagging category.

Two different analysis techniques, full reconstruction~\cite{ref:run1-2-breco}  
and partial reconstruction~\cite{ref:run1-2-ihbd}  were used for 
the $\sin(2\beta+\gamma)$ measurement with $\Bz \rightarrow D^{(*)\mp} \pi^{\pm}$.

The full reconstruction technique is used to measure the \CP asymmetry in 
$\Bz \rightarrow D^{(*)\mp} \pi^{\pm}$ and  $\Bz \rightarrow D^{*\mp} \rho^{\pm}$ 
decays~\cite{ref:full-reco}. From a time-dependent maximum likelihood fit 
the following parameters related to the \CP violation angle 
$2\beta\!+\!\gamma$ are obtained:
\begin{eqnarray} 
  a^{D\pi}              &=& -0.010\pm 0.023  \,\pm 0.007 \,\,  \nonumber \\
  c_{\rm lep}^{D\pi}    &=& -0.033\pm 0.042  \,\pm 0.012 \,\,  \nonumber \\
  a^{D^*\pi}            &=& -0.040\pm 0.023  \,\pm 0.010 \,\, \\
  c_{\rm lep}^{D^*\pi}  &=& \phantom{-}0.049\pm0.042  \,\pm 0.015 \,\, \nonumber \\
  a^{D\rho}             &=& -0.024\pm 0.031  \,\pm 0.009 \,\,  \nonumber \\
  c_{\rm lep}^{D\rho}   &=& -0.098\pm 0.055  \,\pm 0.018 \,\, \nonumber 
\end{eqnarray}
where the first error is statistical and the second is systematic. 
The systematic error for  $\Bz \rightarrow D^{*\mp} \rho^{\pm}$ includes the maximum bias of
asymmetry parameters due to possible dependence of $r$ on the $\pi\pi^0$ invariant mass. 
For the measurement of $2r\cos(2\beta+\gamma)\sin\delta$ parameter only the lepton-tagged events are used due to 
a presence of tag-side \CP violation effect~\cite{ref:abc}.

\begin{figure}[!htb]
  \includegraphics[width=0.49\linewidth]{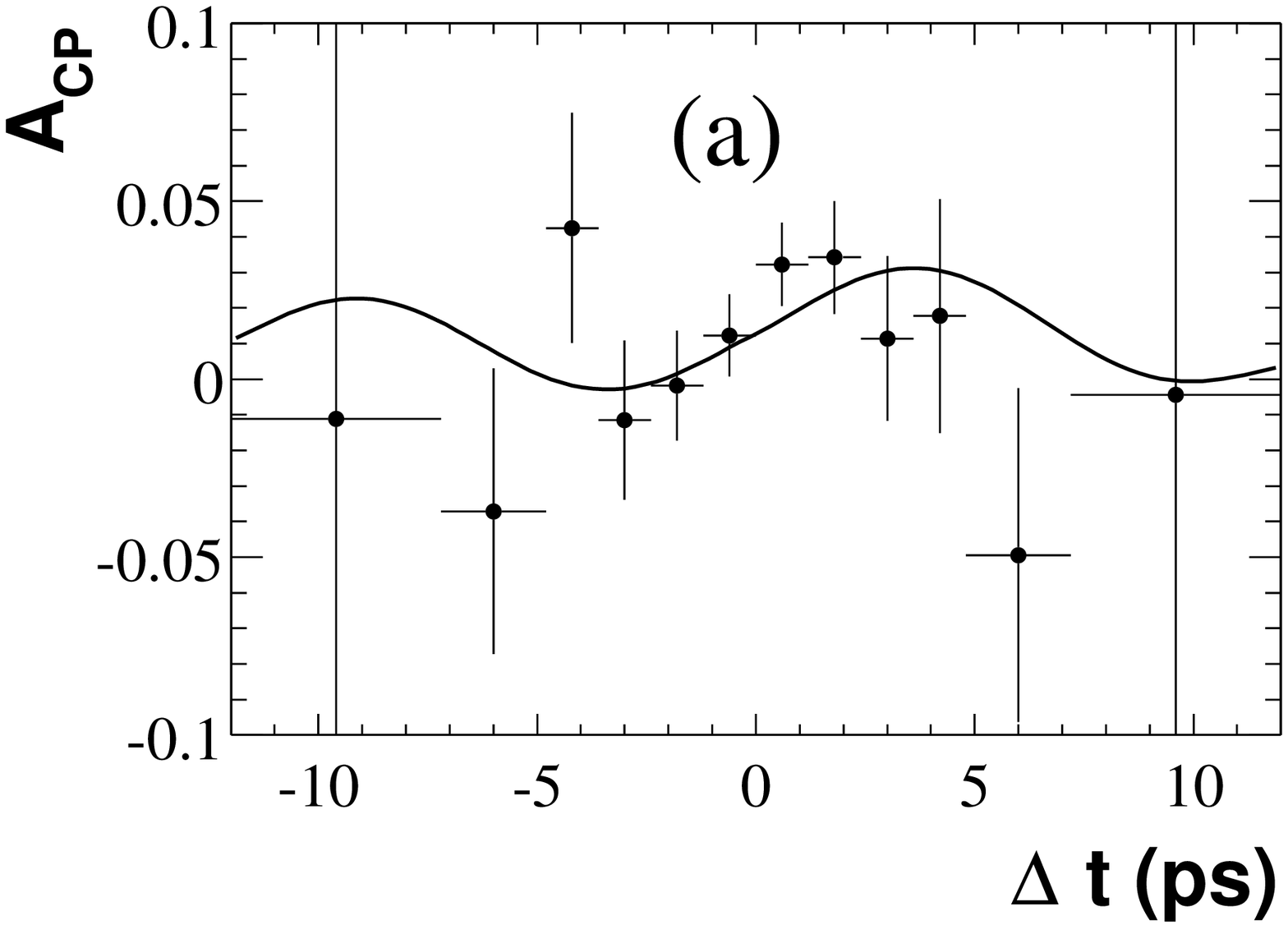}
  \includegraphics[width=0.49\linewidth]{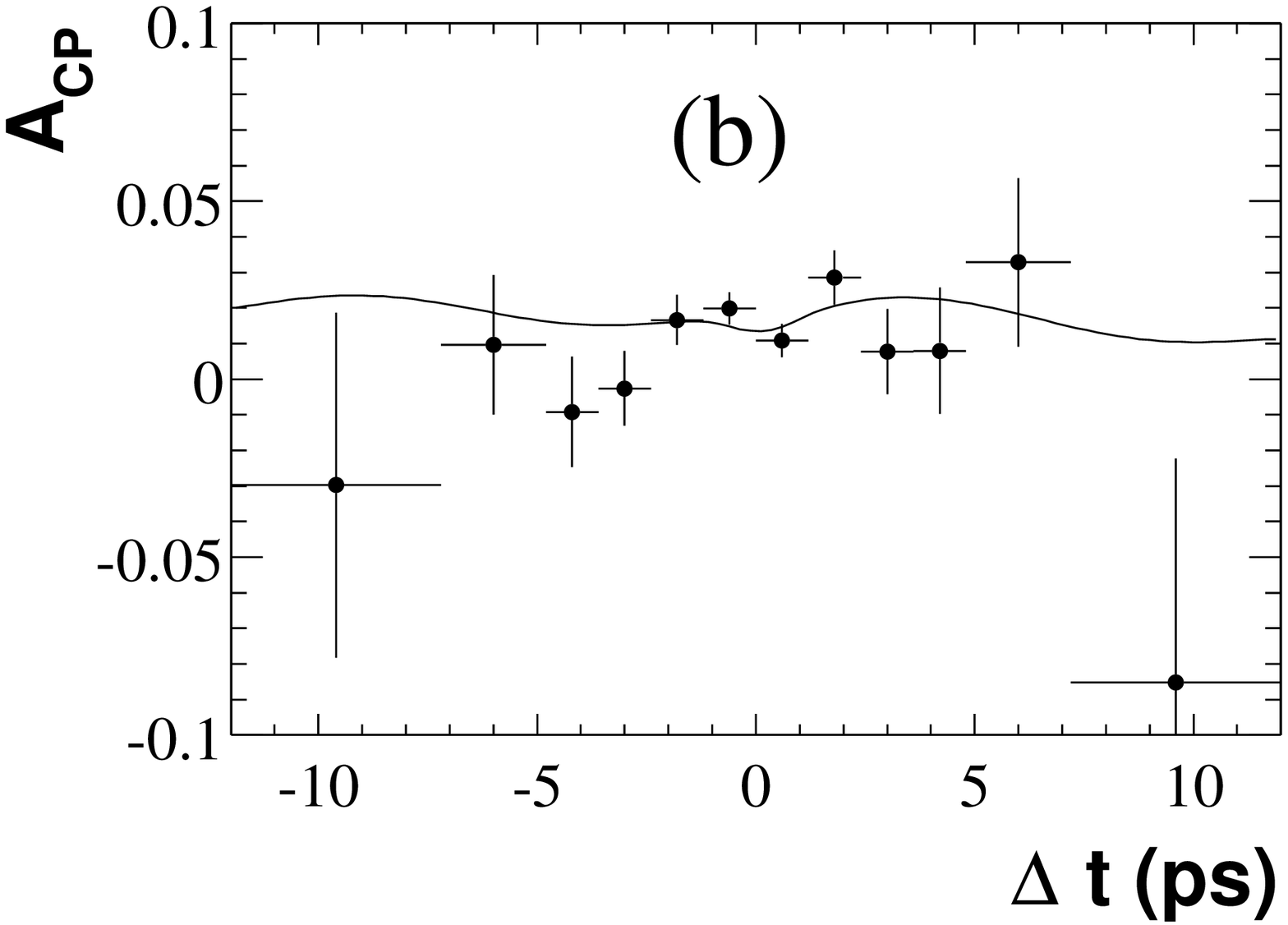}
  \caption{  \label{fig:sin2bg_asym} Raw asymmetry for (a) lepton-tagged and (b) kaon-tagged
    events of $\btodstpipm$ decay mode using the method of the partial reconstruction. 
    The curves represent  the projections of the PDF for the raw
    asymmetry.}
\end{figure}

In the partial reconstruction of a $\btodstpipm$ candidate,
only the hard (high-momentum) pion track $\pi_h$ from the $B$ decay and the
soft (low-momentum) pion track $\pi_s$ from the decay
$D^{*-}\rightarrow \Dzb \pi_s^-$ are used.
Applying kinematic constraints consistent with the signal decay mode,
the four-momentum of the non-reconstructed, ``missing''
$D$ is calculated. Signal events are peaked
in the $\mmiss$ distribution at the nominal $\Dz$ mass.
This method eliminates the efficiency loss associated with the
neutral $D$ meson reconstruction. The \CP asymmetry independent 
on the assumption on $r^{D^*\pi}(r^*)$ measured with this technique is~\cite{ref:part-reco}
\begin{eqnarray}
a^{D^*\pi} &=& -0.034\pm0.014\pm 0.009 \,  \nonumber \\
c_{\rm lep}^{D^*\pi} &=& -0.019\pm0.022\pm 0.013 \,
\label{math:dstpi_partial}
\end{eqnarray}
where the first error is statistical and the second is systematic. 
This measurement deviates from zero by 2.0 standard deviations.
Figure~\ref{fig:sin2bg_asym} shows the raw, time-dependent \CP asymmetry
\begin{equation}
A(\dt) = {N_{\Bz}(\dt) - N_{\Bzb}(\dt)\over
                N_{\Bz}(\dt) + N_{\Bzb}(\dt)}
\end{equation}
In the absence of background and with high statistics, perfect tagging, and
perfect $\dt$ measurement, $A(\dt)$ would be a sinusoidal
oscillation with amplitude $2r\sin(2\beta+\gamma)\cos\delta$.

\begin{figure}[!htb]
  \includegraphics[width=0.48\textwidth]{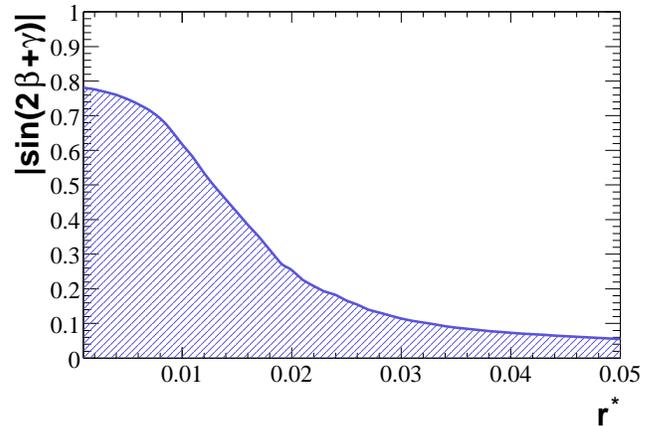}
\caption{\label{fig:limit-vs-r} Lower limit on $|\sin(2\beta+\gamma)|$ a
t 90\% CL as a function of $\r$, for $\r>0.001$.
}
\end{figure}
\begin{figure}[!htb]
  \includegraphics[width=0.48\textwidth]{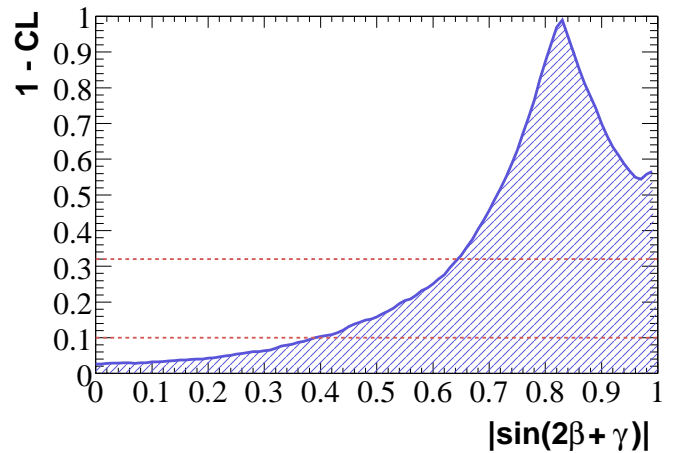}
\caption{\label{fig:s2bgCL} The shaded region denotes the allowed range of $|\sin(2\beta+\gamma)|$
for each confidence level. The horizontal lines show, from top
to bottom, the 68\% and 90\% CL.
}
\end{figure}

Two methods for interpreting these results in terms of
constraints on $|\sin(2\beta+\gamma)|$ are used. 
Both methods involve minimizing a $\chi^2$
function that is symmetric under the exchange $\sin(2\beta+\gamma) \rightarrow
-\sin(2\beta+\gamma)$, and applying the method of Ref.~\cite{ref:Feldman}.
In the first interpretation method, no assumption regarding the value of $r^*$ is made.
The resulting 95\% lower limit for the mode $\Bz\to D^{*\mp}\pi^\pm$ is shown as a function of $r^*$ in
Figure~\ref{fig:limit-vs-r}. 
The second interpretation assumes that $r^{(*)}$ can be estimated from the
Cabibbo angle, the ratio of branching fractions ${\cal
B}(B^0\rightarrow {D^{(*)}}_s^{+} \pi^-) / {\cal B}(B^0\rightarrow
{D^{(*)}}^{-} \pi^+)$, and the ratio of decay constants
$f_{\Dstar} / f_{\Dstar_s}$.
The confidence level as a function of $|\sin(2\beta+\gamma)|$ is shown in Figure~\ref{fig:s2bgCL}.
This method yields the lower limits $|\sin(2\beta+\gamma)|\!>\!0.64\ (0.40)$ at $68\%$ $(90\%)$ C.L.

\section{$\Bzb\to D^{(*)0}\bar{K}^{(*)0}$ decays}
The decay modes $\Bzb\to D^{(*)0}\bar{K}^0$ have been proposed for determination 
of $\sin(2\beta+\gamma)$ from measurement of time-dependent \CP asymmetries~\cite{ref:th-dst0k0}. 
In the Standard Model the decays of \Bz and \Bzb mesons into final state 
$D^{(*)0}\KS$ proceed through the $\b \to \c$ and $\overline{b}\to \overline{u}$ amplitudes, respectively. 
Due to relatively large \CP asymmetry 
($r_B\equiv|A(\Bzb\to \bar{D}^{(*)0}\bar{K}^0)|/|\Bzb\to D^{(*)0}\bar{K}^0)|\simeq 0.4$) 
these decay channels look very  attractive for such a measurement. Since the parameter $r_B$ 
can be measured with sufficient data sample by fitting the $C$ coefficient  in time distributions, 
the measured asymmetry can be interpreted in terms of $\sin(2\beta+\gamma)$ without additional assumptions. 
However, the branching fractions of such decays are relatively small ($\sim5\cdot10^{-5}$). 
That is way the large data sample is still required. 

\begin{figure}[!htb]
  \includegraphics[width=0.49\linewidth]{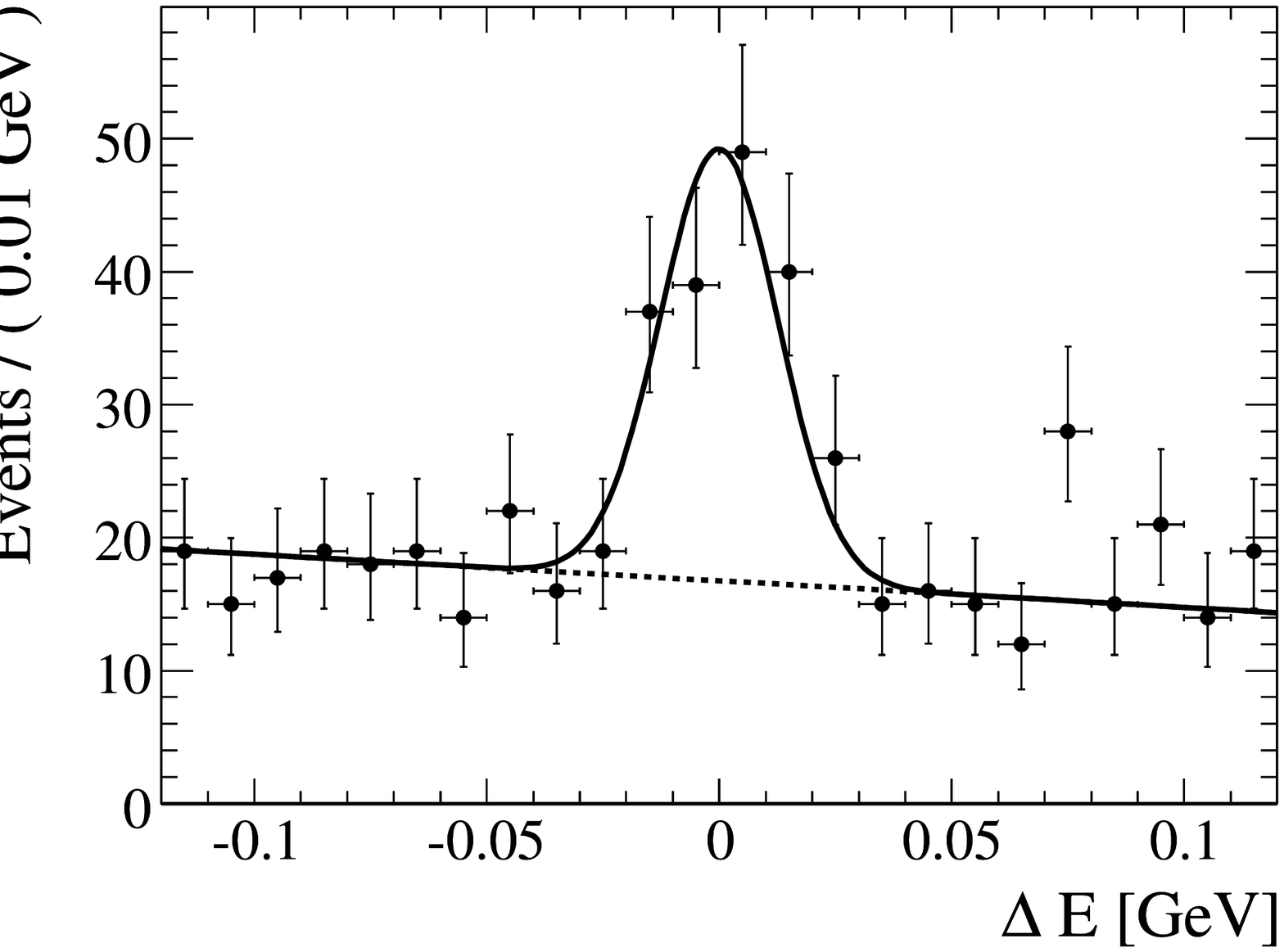}
  \put(-102,74){a)}%
  \includegraphics[width=0.49\linewidth]{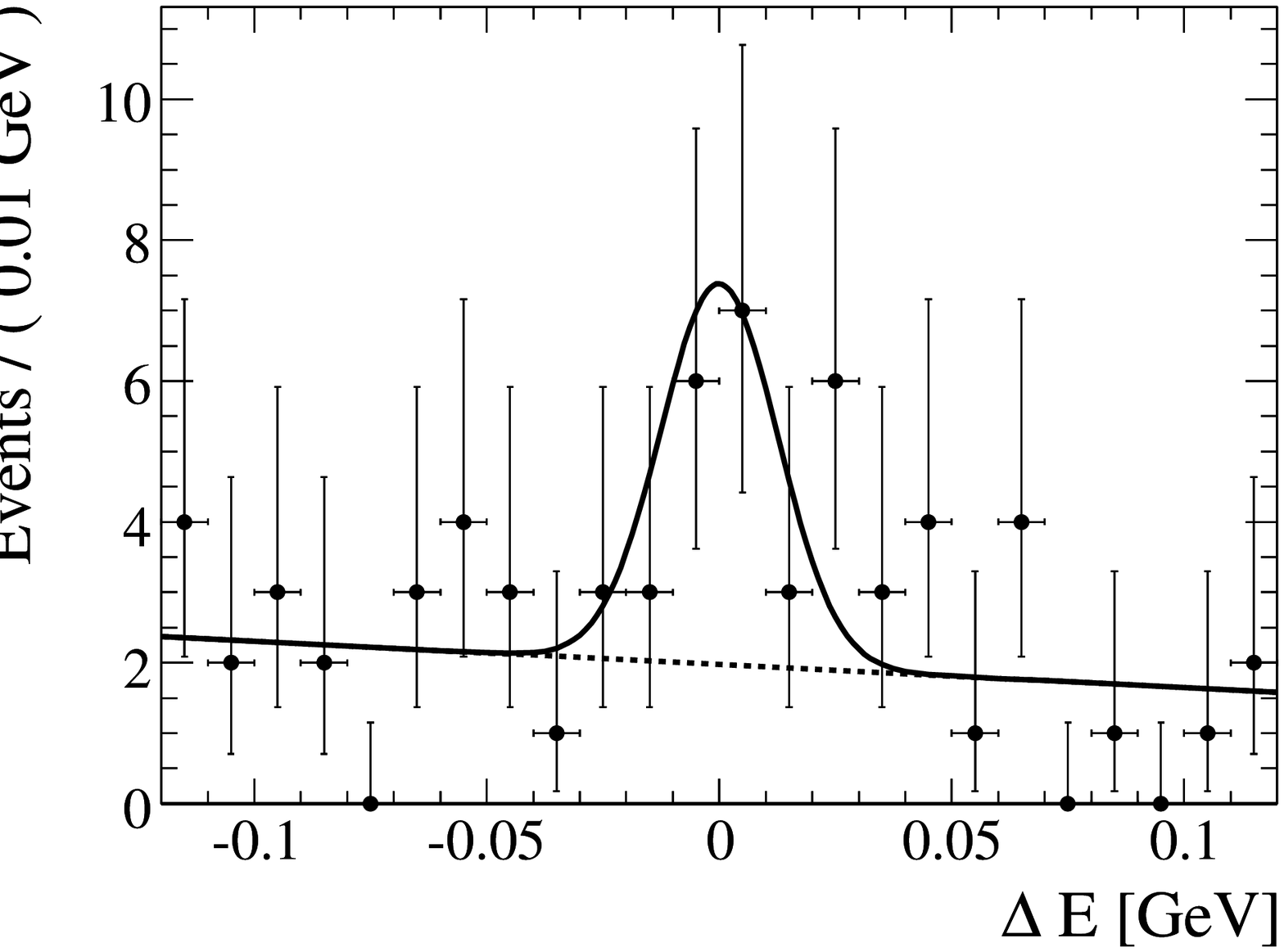}
  \put(-102,74){b)}\\
  \includegraphics[width=0.49\linewidth]{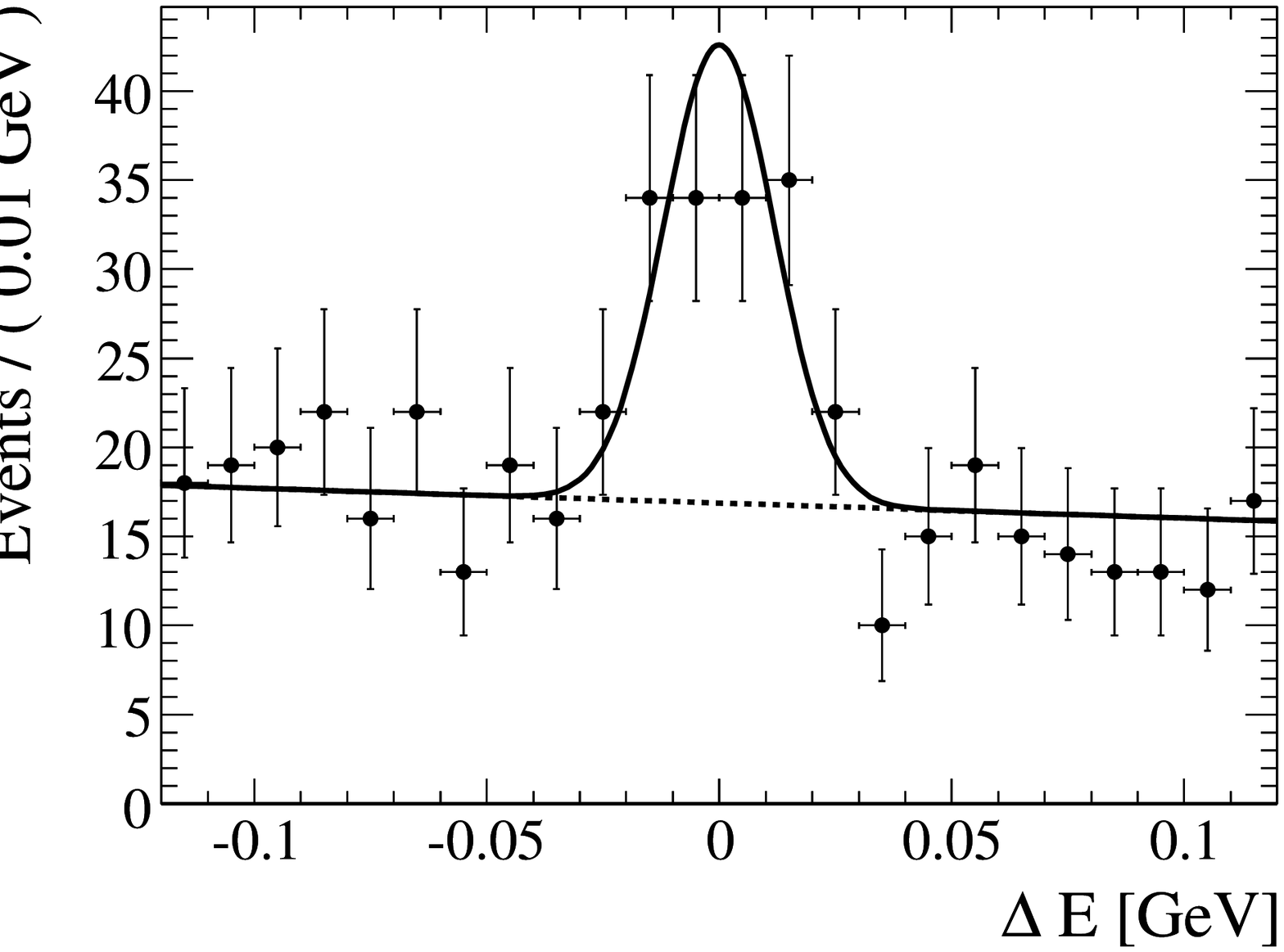}
  \put(-102,74){c)}%
  \includegraphics[width=0.49\linewidth]{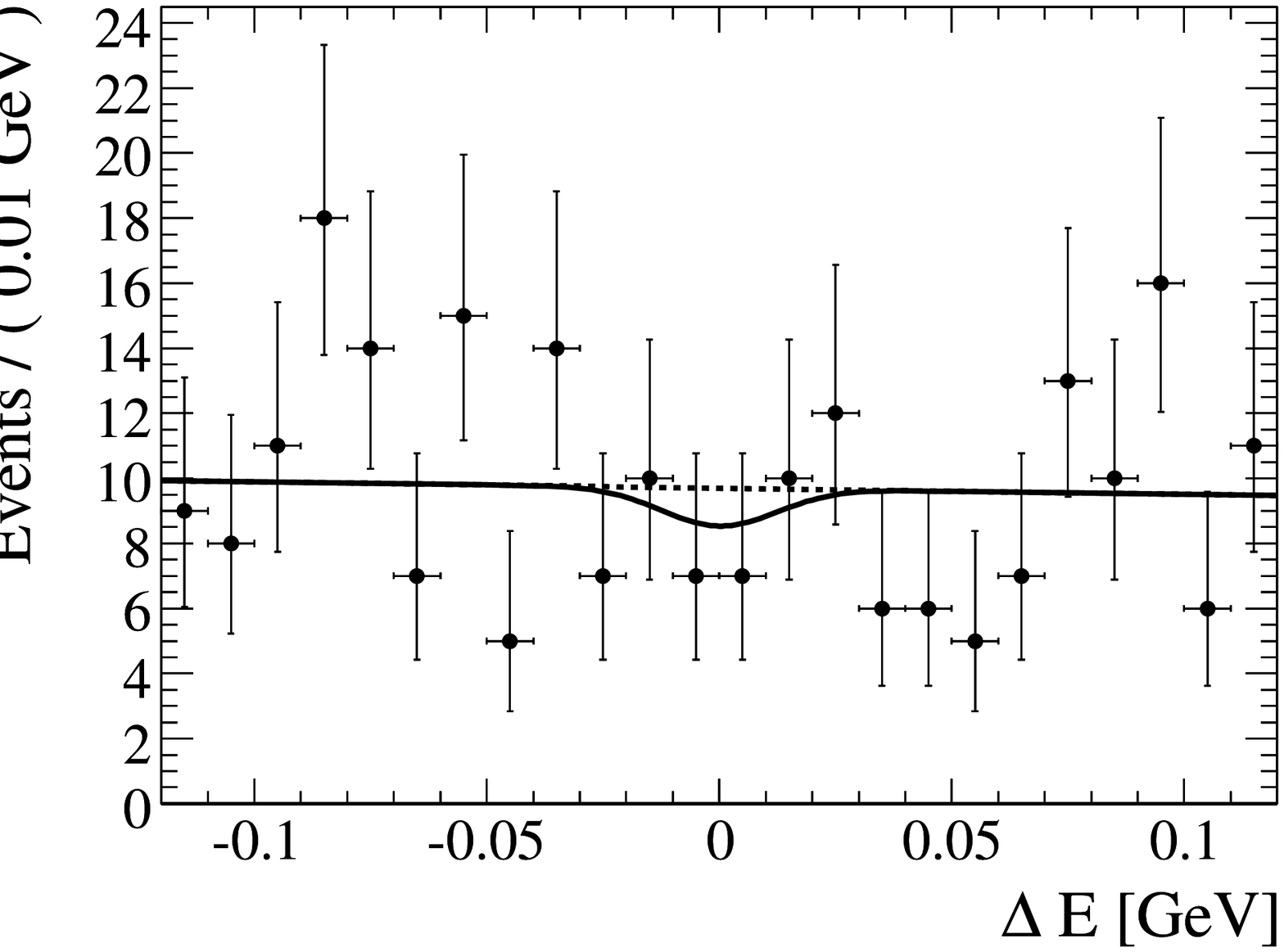}
  \put(-102,74){d)}%
  \caption{\label{fig:DK}
    Distribution of \DeltaE\ for  a) $\Bzb\to \Dz\Kzb$, b)
    $\Bzb\to \Dstarz\Kzb$, c) $\Bzb\to \Dz\Kstarzb$, and d)
    $\Bzb\to \Dzb\Kstarzb$ candidates with $|\mes-5280~\mevcc|< 8\mevcc$. The
    points are the data, the solid curve is the projection of the likelihood fit, and
    the dashed curve represents the background component.
  }
\end{figure}

From the measured signal yields~\cite{ref:dst0kst0}, we find 
\begin{eqnarray}
  \BR(\Bzb\to \Dz\Kzb)     &=& (5.3\pm0.7\pm0.3)\times 10^{-5}\, \nonumber \\
  \BR(\Bzb\to \Dstarz\Kzb) &=& (3.6\pm1.2\pm0.3)\times 10^{-5}\,  \label{eq:brdk} \\
  \BR(\Bzb\to \Dz\Kstarzb) &=& (4.0\pm0.7\pm0.3)\times 10^{-5}\, \nonumber \\
  \BR(\Bzb\to \Dzb\Kstarzb)&<& 1.1\times 10^{-5}\ {\rm at}\ 90\% \ {\rm C.L.} \nonumber
\end{eqnarray}
where the uncertainties are statistical and systematic, respectively. 
Figure~\ref{fig:DK} shows the \DeltaE\ distributions of
candidates with $|\mes-5280|<8$~\mevcc for the
sums of the reconstructed \Dz\ decay modes.

The \B decay dynamics can modify 
the expectation for the ratio $r_B$. The magnitude of this ratio can be probed by measuring 
the rate for the decays  $\Bzb\to D^{(*)0}\bar{K}^{*0}$ and  $\Bzb\to \bar{D}^{(*)0}\bar{K}^{*0}$
using the self-tagging decay $\Kstarzb\ra\Km\pip$. 
The $\Bzb\ra\Dz\Kstarzb$ and $\Bzb\ra\Dzb\Kstarzb$ decays are distinguished by the
correlation between the charges of the kaons produced in the
decays of the neutral $D$ and the \Kstarzb. 
This charge correlation in the final state is diluted by the presence of the
doubly-Cabibbo-suppressed decays
$\Dz\ra\Kp\pim,\Kp\pim\piz$, and $\Kp\pim\pip\pim$.
The ratio $r_B$ is related to the experimental observable ${\mathcal R}$ defined for the $\Dz\ra\Kp\pim$ decay as
\begin{eqnarray}
{\mathcal R} & = & \frac{\Gamma(\Bzb\ra(\Kp \pi^{-})_{D}\Kstarzb)}
                          {\Gamma(\Bzb\ra(\Km \pi^{+})_{D}\Kstarzb)}\nonumber \\
             & = & \r_B^{2} + r_{D}^{2} + 2\r_B r_{D}\cos(\gamma+\delta)\label{eq:R_ADS} ,
\end{eqnarray}
where
\begin{eqnarray}
r_{D} & = & \frac{|{\mathcal A}(\Dz\ra\Kp \pi^{-}  )|}
                 {|{\mathcal A}(\Dz\ra\Km  \pi^{+} )|},\\
 \delta & = & \delta_{B} + \delta_{D},
\end{eqnarray}
and $\delta_{B}$ and $\delta_{D}$ are strong phase differences between the two $B$ and
$D$ decay amplitudes, respectively. 
From the measured \B branching fractions (Eq.~\ref{eq:brdk}), values of $r_D$~\cite{bib:PDG2004} and Eq.~\ref{eq:R_ADS}, 
one obtains $r<0.40$ at the 90\% C.L. To conclude, the present signal yields combined with this limit on $r$ suggest that
a substantially larger data sample is needed for a competitive
time-dependent measurement of $\sin(2\beta+\gamma)$ in $\Bzb\ra D^{(*)0}\Kzb$ decays.

\section{$\Bz \rightarrow D^{(*)\mp} a_{0(2)}^{\pm}$ decays}
Recently it was proposed to consider the $\Bz \rightarrow D^{(*)\mp} a_{0(2)}^{\pm}$ decays 
for measurement of $\sin(2\beta+\gamma)$~\cite{ref:th-dsta0_a2}.
The decay amplitudes of \B mesons to light scalar or tensor mesons such as $a_0^+$ or $a_2^+$, 
emitted from a weak current, are significantly suppressed  due to the small decay constants $f_{a_{0(2)}}$. 
Thus, the absolute value of the CKM-suppressed and favored amplitudes become comparable. As a result, the 
\CP asymmetry in such decays is expected to be large. However, the theoretical predictions of the branching 
fractions for  $\Bz \rightarrow D^{(*)\mp} a_{0(2)}^{\pm}$ is expected of the order of 
$(1\div4)\cdot 10^{-6}$~\cite{ref:br-dsta0_a2}. The main uncertainty in the branching fractions 
of these decay modes is due to unknown $\B\to a_{0(2)}X$ transition form factors. 
One way to verify the expectations and test a validity of the factorization approach is to measure the branching 
fractions for the more abundant decay modes $\Bz\to D_s^{(*)+}a_{0(2)}$. 

Using a sample of about 230 million $\FourS\to\BB$ no evidence for these decays were observed~\cite{ref:dsa}. 
This allowed one to set  upper limits at 90\% C.L. on the branching fractions to be 

\begin{eqnarray}
\BR(\Bz\to D^{+}_s a_{0}^-)&<& 1.9\cdot 10^{-5} \,   \nonumber \\
\BR(\Bz\to D^{+}_s a_{2}^-)&<& 1.9 \cdot 10^{-4}\,   \\
\BR(\Bz\to D^{*+}_s a_{0}^-)&<& 3.6\cdot 10^{-5}\,   \nonumber \\
\BR(\Bz\to D^{*+}_s a_{2}^-)&<& 2.0 \cdot 10^{-4}\,  \nonumber 
\end{eqnarray}

\begin{figure}
  \includegraphics[width=0.49\textwidth]{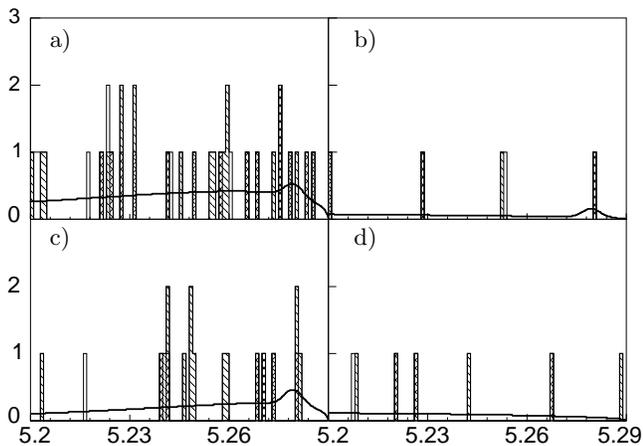}
    \put(-230,150){a)}
    \put(-230,75){c)}
    \put(-115,150){b)}
    \put(-115,75){d)}
\caption{\label{fig:dsa0}      
  Distributions of \mes\ for a) $B^0\to D_s^{+} a_{0}^-$, 
  b) $B^0\to D_s^{+} a_{2}^-$, c) $B^0\to D_s^{*+} a_{0}^-$, 
  d) $B^0\to D_s^{*+} a_{2}^-$ candidates
  overlaid with the projection of the maximum likelihood
  fit. Contributions from the three $D^+_s$ decay modes
  are shown with different hatching styles: $\phi \pi^+$ is cross hatched,
  $\Kstarzb K^+$ is hatched, and $\KS K^+$ is white.}
\end{figure}

Figure~\ref{fig:dsa0} shows the \mes
distributions for the reconstructed candidates 
$B^0 \to D_s^{+} a_0^-$, $B^0 \to D_s^{+} a_2^-$,
$B^0 \to D_s^{*+} a_0^-$ and  $B^0\to
D_s^{*+}a_2^-$. For each \B decay mode,  an
unbinned maximum-likelihood fit is performed
using the candidates from the three \Ds decay modes.

The upper limit value for
$B^0\to D_s^{+} a_0^-$ is lower than the theoretical expectation,
which might indicate the need to revisit the $B \ra a_0 X$ transition
form factor estimate. It might also imply the limited applicability
of the factorization approach for this decay mode. 
The measured upper limits suggest that the branching ratios of $B^0 \ra D^{(*)+} a_{0(2)}^-$
are too small for \CP-asymmetry measurements given the present
statistics of the $B$-factories. 
The measurement of $\sin(2\beta+\gamma)$ in $B^0 \ra D^{(*)+} a_{0(2)}^-$ 
decays is an interesting program for the future experiments such as Super\B-factories.

\section*{Conclusion}
The substantial constraint on the CKM angles $2\beta+\gamma$ comes from  
the measurements of time-dependent \CP asymmetry in the $\Bz{\to}D^{(*)\mp}\pi^{\pm}$ and $\Bz{\to}D^{\mp}\rho^{\pm}$ decays.
The \babar\ experiment has used two techniques such as full and partial reconstruction to increase the signal yields in the 
$D^{*\mp}\pi^\pm$  channel. The combined \babar and {\it BELLE} results~\cite{ref:belle-dstpi} 
for \CP violation in the most precisely measured decay channel $D^{*\mp}\pi^\pm$ is
\begin{equation}
a^{D^*\pi} = 2 r^* \sin(2\beta+\gamma)\cos\delta = -0.037\pm 0.011\, 
\end{equation}
This measurement performed at the level of one per cent deviates from zero by 3.4 standard deviations. 
Future updates are therefore of a great interest. 
We interpret the \babar\ result in terms of $\sin(2\beta+\gamma)$ and find $|\sin(2\beta+\gamma)|\!>\!0.64$ $(0.40)$ at $68\%$ $(90\%)$ C.L.
using a frequentist method.

The \babar\ experiment has measured the branching fractions of $\Bzb\to D^{(*)0}\bar{K}^{(*)0}$ 
and set up the limit on $\Bz \rightarrow D^{(*)\mp} a_{0(2)}^{\pm}$ decays. 
The present signal yields and established limits suggest that
a substantially larger data sample is needed for a competitive
time-dependent measurement of $\sin(2\beta+\gamma)$ with these decay channels.


\end{document}